\documentclass{article}
\usepackage{emulateapj,apjfonts}
\input psfig

\newcommand{\etal}{{\it et~al.}}
\newcommand{\eg}{{\it e.g.}}
\newcommand{\ie}{{\it i.e.}}
\newcommand{\Msun}{M_\odot}

\newcommand{\kms}{$\rm {km}~\rm s^{-1}$}
\newcommand{\mtl}{{\it M/L}$_{\rm V}$}
\newcommand{\vdm}{van~der~Marel}
\newcommand{\degr}{$^\circ$}

\begin{document}

\lefthead{NGC~3379}
\righthead{Gebhardt~\etal}
\submitted{AJ}

\title{Axisymmetric, 3-Integral Models of Galaxies: A Massive Black
Hole in NGC~3379}

\author{Karl Gebhardt\altaffilmark{1,2} and Douglas Richstone}
\affil{Dept. of Astronomy, Dennison Bldg., Univ. of Michigan, Ann
Arbor 48109} \affil{gebhardt@ucolick.org, dor@astro.lsa.umich.edu}

\altaffiltext{1}{Hubble Fellow} \altaffiltext{2}{Current Address:
UCO/Lick Observatory, University of California, Santa Cruz, CA 95064}
 
\author{John Kormendy}
\affil{Institute for Astronomy, University of Hawaii,
2680 Woodlawn Dr., Honolulu, HI 96822}
\affil{kormendy@oort.ifa.hawaii.edu}
 
\author{Tod R. Lauer}
\affil{Kitt Peak National Observatory,
National Optical Astronomy Observatories, P. O. Box 26732, Tucson, AZ
85726}
\affil{lauer@noao.edu}

\author{Edward A. Ajhar}
\affil{Kitt Peak National Observatory,
National Optical Astronomy Observatories, P. O. Box 26732, Tucson, AZ
85726}
\affil{ajhar@noao.edu}

\author{Ralf Bender}
\affil{Universit\"ats-Sternwarte, Scheinerstra\ss e1, M\"unchen 81679,
Germany}
\affil{bender@usm.uni-muenchen.de}

\author{Alan Dressler}
\affil{The Observatories of the Carnegie Institution of Washington, 813
Santa Barbara St., Pasadena, CA 91101}
\affil{dressler@ociw.edu}
 
\author{S. M. Faber}
\affil{UCO/Lick Observatories, Board of Studies in Astronomy and
Astrophysics, University of California, Santa Cruz, CA 95064}
\affil{faber@ucolick.org}

\author{Carl Grillmair} 
\affil{Jet Propulsion Laboratory, Mail Stop 183-900, Caltech, 4800 Oak
Grove Dr., Pasadena, CA 91109}
\affil{carl@grandpa.jpl.nasa.gov}

\author{John Magorrian} 
\affil{Institute of Astronomy, Madingley Road, Cambridge CB3 0HA, England} 
\affil{magog@ast.cam.ac.uk}

\author{Scott Tremaine} 
\affil{Princeton University Observatory, Peyton Hall, Princeton, 
NJ 08544} 
\affil{tremaine@astro.princeton.edu}
 
\begin{abstract}

We fit axisymmetric 3-integral dynamical models to NGC~3379 using the
line-of-sight velocity distribution obtained from HST/FOS spectra of
the galaxy center and ground-based long-slit spectroscopy along four
position angles, with the light distribution constrained by WFPC2 and
ground-based images. We have fitted models with inclinations from
29\degr\ (intrinsic galaxy type E5) to 90\degr\ (intrinsic E1) and
black hole masses from 0 to $10^9\Msun$.

The best-fit black hole masses range from
$6\times10^7-2\times10^8\Msun$, depending on inclination. The
preferred inclination is 90\degr\ (edge-on); however, the constraints
on allowed inclination are not very strong due to our assumption of
constant \mtl. The velocity ellipsoid of the best model is not
consistent with either isotropy or a two-integral distribution
function. Along the major axis, the velocity ellipsoid becomes
tangential at the innermost bin, radial in the mid-range radii, and
tangential again at the outermost bins. The rotation rises quickly at
small radii due to the presence of the black hole. For the acceptable
models, the radial to tangential (($\sigma_\theta^2+\sigma_\phi^2)/2)$
dispersion in the mid-range radii ranges from $1.1 < \sigma_r/\sigma_t
< 1.7$, with the smaller black holes requiring larger radial
anisotropy. Compared with these 3-integral models, 2-integral
isotropic models overestimate the black hole mass since they cannot
provide adequate radial motion. However, the models presented in this
paper still contain restrictive assumptions---namely assumptions of
constant \mtl\ and spheroidal symmetry---requiring yet more models to
study black hole properties in complete generality.

\end{abstract}
 
\section{Introduction}

The dynamical state and mass distribution in the central regions of
elliptical galaxies provide clues to the formation and evolutionary
histories of these galaxies (see Merritt 1999 for a
review). Consequently, the existence of central black holes has been
the target of intense scrutiny (see Kormendy~\& Richstone 1995 for a
review). In this paper, we study the elliptical galaxy NGC~3379 using
both {\it Hubble Space Telescope} and ground-based photometric and
kinematic data in order to understand its central dynamics.

At ground-based resolution NGC~3379 is a prototypical elliptical (E1)
galaxy with M$_V=-20.6$ (Faber~\etal\ 1997) at a distance of 10.4~Mpc
(Ajhar~\etal\ 1997). Lauer~\etal\ (1995) classify it as a ``core''
galaxy---a galaxy that has a break in the surface brightness profile
but still maintains a rising profile into the smallest measured
radius. Faber~\etal\ (1997) hypothesize that core galaxies are
associated with massive black holes. However, few core galaxies have
strong black hole detections: M87 (Harms~\etal\ 1994), M84
(Bower~\etal\ 1998), and N4261 (Ferrarese~\etal\ 1996). These
detections result from gas dynamics, not stellar dynamics. Stellar
dynamical evidence for central black holes in core galaxies is
difficult to obtain since (1) core galaxies have low surface
brightnesses, making kinematic observations difficult, (2) the lack of
rotation complicates the dynamical modeling since velocity
anisotropies potentially govern the dynamical support, and (3) stars
near the center travel out to the core radius so their contribution to
the central velocity profile is de-weighted (Kormendy 1992). NGC~3379
was chosen for this study because of its relatively high central
surface brightness among core galaxies, and because of its otherwise
normal morphological and dynamical structure.

The proximity, brightness, and normalcy of NGC~3379 make it the object
of many ground-based photometric (de~Vaucouleurs \& Capaccioli 1979,
Lauer 1985, Capaccioli~\etal\ 1987, Capaccioli~\etal\ 1991) and
kinematic (Kormendy 1997, Kormendy 1985, \vdm\ \etal\ 1990,
Bender~\etal\ 1994, Statler 1994, Statler~\& Smecker-Hane 1999)
studies.  However, previous models for NGC~3379 have yielded
inconsistent results. Assuming that the departure of the light profile
from the $R^{1/4}$ law signifies the existence of a stellar disk,
Capaccioli~\etal\ (1991) find that the most likely model is nearly
face-on with an inclination of 31\degr\ (intrinsic E5 galaxy). Their
conclusion was that NGC~3379 and NGC~3115 have the same intrinsic
shape but are seen from different viewing angles. In their model for
NGC~3379 the central region is nearly oblate, but at about $r=0.5 r_e$
($r_e=35$\arcsec) the model becomes triaxial. No dynamical information
was used in this model. A major uncertainty is that slight departures
from a $R^{1/4}$ law, on which their model is based, may be due to
tidal effects (Kormendy 1977) instead of a disk. Furthermore, there is
no reason to expect galaxies to follow a $R^{1/4}$ law exactly.

Kinematic information is required to determine unambiguously the
inclination of a spheroidal system. Van~der~Marel \etal\ (1990), using
the ground-based velocity data of Davies \& Birkinshaw (1988) and
Franx~\etal\ (1989), conclude that the inclination of NGC~3379 is
60\degr; however the two datasets are not consistent and give slightly
different velocity anisotropy parameters. Van~der~Marel \etal\ used
2-integral flattened models with a parametrized form for the
anisotropy. No goodness-of-fit criteria were given, and thus we are
not able to compare their results with those of others. Statler
(1994), using Bayesian statistics on the same datasets, prefers
models that have inclination greater than 45\degr. Statler rejects
Capaccioli's conclusion that the galaxy is flattened and triaxial at
the 98\% confidence level, but his results could accommodate those of
\vdm\ \etal\ Statler uses only the mean velocity measured along four
position angles; the radial variation in the velocity or the
dispersion are not considered. Furthermore, Statler uses data only
outside 15\arcsec. However, Statler has obtained new, high S/N data
(Statler~\& Smecker-Hane 1999), which he will present in a future
paper.

The studies above use low S/N, sparse kinematic data, or no kinematic
data at all. In this paper, we report results from observations of the
full line-of-sight velocity distribution (LOSVD) at various position
angles from high S/N ground-based data and data from HST. The data are
fit to axisymmetric, three-integral orbit-based models.
Richstone~\etal\ (1999) describe the modeling technique in a companion
paper. We exploit the full LOSVDs {\it directly} in the
modeling. Alternatively, we could use the moments of the LOSVDs;
however, the full shape of the LOSVD contains important information on
the underlying dynamics, and it is desirable to incorporate the full
LOSVD when the spectra have sufficient signal-to-noise to determine
this reliably. In fact, our result for NGC~3379 critically depends on
the shape of the central LOSVD. Cretton~\etal\ (1999) and \vdm~\etal\
(1998) present a very similar modeling technique, where the only
difference from ours is that we fit the full LOSVD as opposed to their
moment fitting procedure. There are other programs that also use the
higher moments of the LOSVD as constraints (Dejonghe, 1987; Gerhard
1993; Rix~\etal\ 1997). Obtaining the full LOSVD at many position
angles is the maximum obtainable kinematic information and has now
become routine (\eg\ Bender~\etal\ 1994, Carollo~\etal\ 1995, Gebhardt
\& Richstone 1999).

In \S2, we describe the photometric and kinematic data for
NGC~3379. In \S3, we describe the models and present results in
\S4. Discussion is given in \S5.

\section{Data}

\subsection{Photometry: HST and Ground-Based Imaging}

We obtained Hubble Space Telescope (HST) observations of NGC~3379,
consisting of four 500s F555W ($V$-band) and four 400s F814W
($I$-band) WFPC2 images, on 1994 November 19. Centering the galaxy in
the high-resolution PC1 camera (0.0455\arcsec\ pixel$^{-1}$) provided
a total signal in the central pixel of $\sim 2\times 10^4 e^-$ in both
$V$ and $I$. We deconvolved the summed images with 40 Lucy-Richardson
iterations. The surface brightness, ellipticity, and position angle of
the major axis were obtained using the same procedure as in
Lauer~\etal\ (1995) and are presented in Table~1.

\vskip 0.2cm \psfig{file=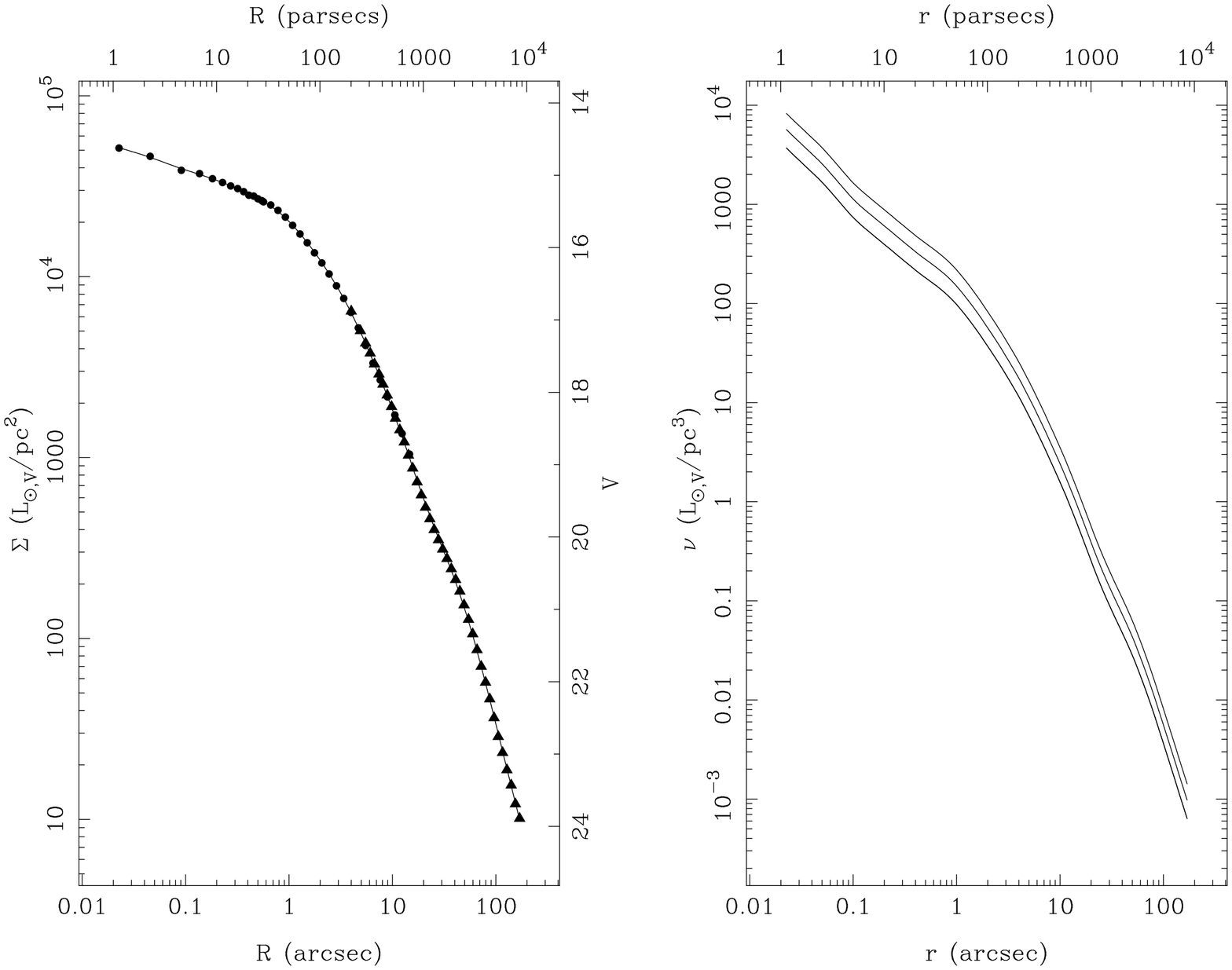,width=9.1cm,angle=-90}
\figcaption[n3379sd.ps]{Visual surface brightness and luminosity
density for NGC~3379. The solid circles are the data from HST and the
solid triangles are the data from Peletier~\etal\ (1990), which have
been adjusted to match with the HST data assuming constant color. The
solid line in the left panel represents a smoothing spline
(Gebhardt~\etal\ 1996) of the surface brightness data. The inferred
luminosity density is given for three different inclinations; from
bottom to top the lines represent the densities at 90\degr\ (intrinsic
E1), 31\degr\ (E4), and 27\degr\ (E6).\label{fig1}}
\vskip 0.3cm

The HST profile reaches only about 12\arcsec\ from the center and must
be extended to larger radii using ground-based
photometry. Peletier~\etal\ (1990) have obtained ground-based $UBR$
photometry for NGC~3379 over the radius range 4\arcsec--170\arcsec,
and Capaccioli~\etal\ (1990) have $B$ photometry out to 500\arcsec; we
must therefore estimate the transformations to $V$ and $I$ using the
overlap region with the HST data. The data from Peletier~\etal\ and
Capaccioli~\etal\ agree well, and since Peletier~\etal\ provide
several colors, we use their results. The colors in the overlap
region, based on the average differences there, are $B-V=-0.58$ and
$R-I=-0.94$. We have ignored possible variations in the colors as a
function of radius. Using these colors, we combine the HST and
ground-based data to provide the full profile. The left panel in
Fig.~1 shows the surface brightness profile in $V$ along the major
axis from 0.02\arcsec--170\arcsec. The three panels in Fig.~2 show the
ellipticity, the position angle of the major axis (measured from north
through east), and the color $V-I$ as functions of radius along the
major axis. The random uncertainties can be estimated from the local
scatter.

The dynamical modeling described below assumes axisymmetry.  We must
therefore determine the best axisymmetric density distribution
consistent with the observed surface brightness distribution.
Strictly speaking, there is none, because, as shown in Figure~2, there
are abrupt changes in the ellipticity and position angle at radii less
than 1\arcsec \ and larger than 100\arcsec. PA changes in particular
are normally attributed to variation in the axis ratios with radius in
a triaxial distribution, or to real three-dimensional twists in
locally axisymmetric objects in which surfaces of constant density,
while axisymmetric, are not co-axial.

\vskip 0.2cm
\psfig{file=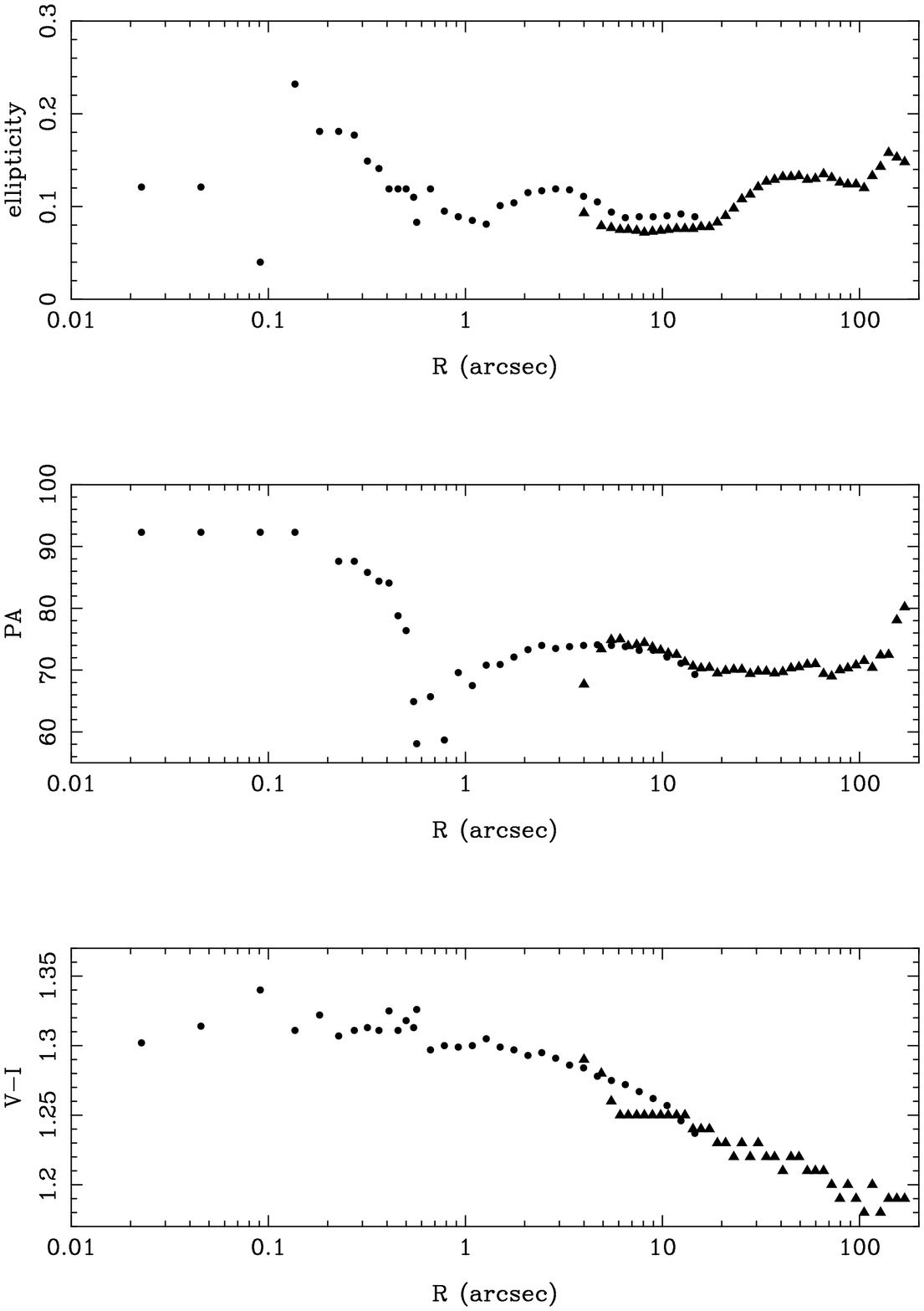,width=9.1cm,angle=0}
\figcaption[n3379pec.ps]{Ellipticity, position angle and $V-I$ color
 as a function of
radius along the major axis for NGC~3379. The solid circles are the
data from HST, and the solid triangles are data from Peletier~\etal\
(1990). The ground-based $V-I$ is estimated from $B-R$.\label{fig2}}
\vskip 0.3cm

However, for several reasons we believe that, at radii less than
100\arcsec, an axisymmetric model will suffice for our purposes.
First, in the middle range of radii, $1\arcsec<r<100\arcsec$, the
ellipticity and PA are fairly constant at 0.1 and 72 $\pm$ 5\degr.
Second, in the central region there is a dust ring with radius
1\arcsec\ (discussed below and seen in Fig.~5).  The absorption by the
dust ring -- 1\% of the galaxy surface brightness -- is enough to
affect estimates of both the ellipticity and the position angle.
Furthermore, the shallow central slope (dlog~$\Sigma$/dlog~$R = -0.2$)
of the inner profile and the small ellipticity ($\sim 0.1$) render the
position angle at small radii ($R<1\arcsec$) highly uncertain. Even if
the PA gradient seen in Fig~2 is real and well determined, the amount
of light, and therefore the gravitational impact, of the associated
departure from axisymmetry is small.  Finally, at large radii,
departures from axisymmetry have little impact on the central
gravitational field.

The data thus suggest that NGC~3379 can be reasonably well represented
by a spheroid of constant ellipticity. The deprojection of the surface
brightness is then unique for a given inclination.  The deprojection
is based on a non-parametric estimate of the density using smoothing
splines (Gebhardt~\etal\ 1996). The right panel of Fig.~1 plots the
luminosity density for three different inclinations (90\degr\ is
edge-on).  For different inclinations, and hence different
flattenings, the luminosity density scales as the ratio of the
apparent to the true axis ratio.

We did not examine models in which the galaxy is assumed to be
axisymmetric but not spheroidal (\ie, with varying ellipticity), in
part because in this case the deprojection is not unique. Romanowsky
\& Kochanek (1997) determine the uncertainties in the deprojection of
an axisymmetric system for various inclinations, and the uncertainties
can be quite large for low inclinations (close to face-on). We could
test the effect of our spheroidal assumption by modeling the most
extreme situation: a galaxy which is seen nearly face-on and
containing an embedded disk (similar to the suggestion of
Capaccioli~\etal\ 1991). The most likely effect is to change to
inclination and \mtl\ constraints, and relaxing the spheroidal
assumption should have little effect on the measured black hole
mass. This issue, however, will not be addressed here.

\subsection{{\it HST} FOS Spectroscopy}

NGC 3379 was observed on 1995, January 23 -- 27 with the 0.21\arcsec\
square aperture (``{\tt 0.25-PAIR}'') of the FOS. The spectrum was
taken in two visits; the total integration time was 215.5 minutes.

The wavelength range, 4566--6815 \AA, includes the Mg~I~{\it b} lines
at $\lambda \simeq 5175$ \AA\ and the Na I D lines at \hbox{$\lambda
\simeq 5892$ \AA.} The spectrum was electronically quarter-stepped,
giving 2064 1/4-diode pixels.  The reciprocal dispersion was 1.09~\AA\
pixel$^{-1}$.  The instrumental velocity dispersion was measured to be
$\sigma_{\rm instr} = $FWHM/2.35$ = 1.76 \pm 0.03$ pixels = $101\pm
2$~\kms\ (internal error). This width is intrinsic to the instrument
and is not strongly affected by how the aperture is illuminated
(Keyes~\etal\ 1995). We therefore make no aperture illumination
corrections to the measured velocity dispersion.

Flat fielding and correction for geomagnetically induced motions (GIM)
were done as in Kormendy~\etal\ (1996).  The flat-field image was
taken with the same {\tt 0.25-PAIR} aperture used for NGC~3379.  The
galaxy exposure was divided into 3 and 4 subintegrations for the two
visits. Because GIM shifts between subintegrations are small, unlike
the large shifts between galaxy exposures and the flat field, we first
averaged all the exposures and used cross-correlation to determine the
shift between this average and the flat field.  The measured shift was
2.27 pixel.  Each subintegration was multiplied by the flat-field
frame shifted by the above amount.  One dead diode was corrected by
dividing the affected pixel intensities by 0.8.  Then GIM shifts
between subintegrations were determined; most shifts were $<0.5$
pixels, with the largest being 0.8 pixels.  Each subintegration was
then shifted to agree with the subintegration taken closest in time to
the comparison spectrum exposure.  Finally, the spectra were added,
weighted by the exposure times.

\subsubsection{Extracting the LOSVD}

Obtaining the internal kinematic information requires a deconvolution
of the observed galaxy spectrum with a representative set of template
stellar spectra. We deconvolve the spectrum using a maximum-penalized
likelihood (MPL) estimate that obtains a non-parametric line-of-sight
velocity distribution (LOSVD). The fitting technique proceeds as
follows: we choose an initial velocity profile in bins. We convolve
this profile with a weighted-averaged template (discussed below) and
calculate the residuals to the galaxy spectrum. The program varies the
velocity profile parameters---bin heights---and the template weights
to provide the best match to the galaxy spectrum. The MPL technique is
similar to that used in Saha~\& Williams (1994), and Merritt
(1997). However, in constrast to these authors we simultaneously fit
the velocity profile and template weights. We impose smoothness on the
LOSVD by the addition to the $\chi^2$ of a penalty function, which is
the integrated squared second derivative (details are in Merritt 1997
and Gebhardt~\& Richstone 1999). We estimate the best smoothing from
bootstrap simulations (described below). Two templates, DSA107-442 (a
field K0III star) and N188-I69 (a K2III star in NGC~188), are used,
both observed with the FOS. Star N188-I69 is the preferred template,
as the fitting routine gives essentially all of the weight to this
star. However, the LOSVD determined using only the other star is
similar, suggesting that the measurement of the LOSVD is robust. The
results from this maximum likelihood technique were compared to
results of a Fourier correlation technique (Bender~1990), and no
significant differences were found.

The LOSVDs obtained in the two visits by HST agree to within their
uncertainties, and also the pointings from the two visits agree to
within 0.01\arcsec. Therefore, we sum the two spectra weighted by
exposure time to derive a combined LOSVD. A parameterization of the
LOSVD by Gauss-Hermite moments (\vdm\ \& Franx 1993, Gerhard 1993)
gives for the first four moments $v=0\pm14$~\kms,
$\sigma=289\pm13$~\kms, $h_3=0.11\pm0.04$, and $h_4=0.04\pm0.04$. The
velocity zero-point is arbitrary and will be discussed in \S4. All
error bars correspond to the 68\% confidence band.

In the lower part of Figure 3 we plot the template spectrum; the noisy
upper line is the averaged FOS spectrum for NGC~3379 (shifted by its
redshift), and the smooth upper line is the template spectrum
convolved with the derived LOSVD. The continuum has been divided out
of both template and galaxy spectra. The spectrum shown here is a
region around the Mg~I~b triplet (5175~\AA) and not the full spectrum
obtained by the FOS.

\vskip 0.2cm \psfig{file=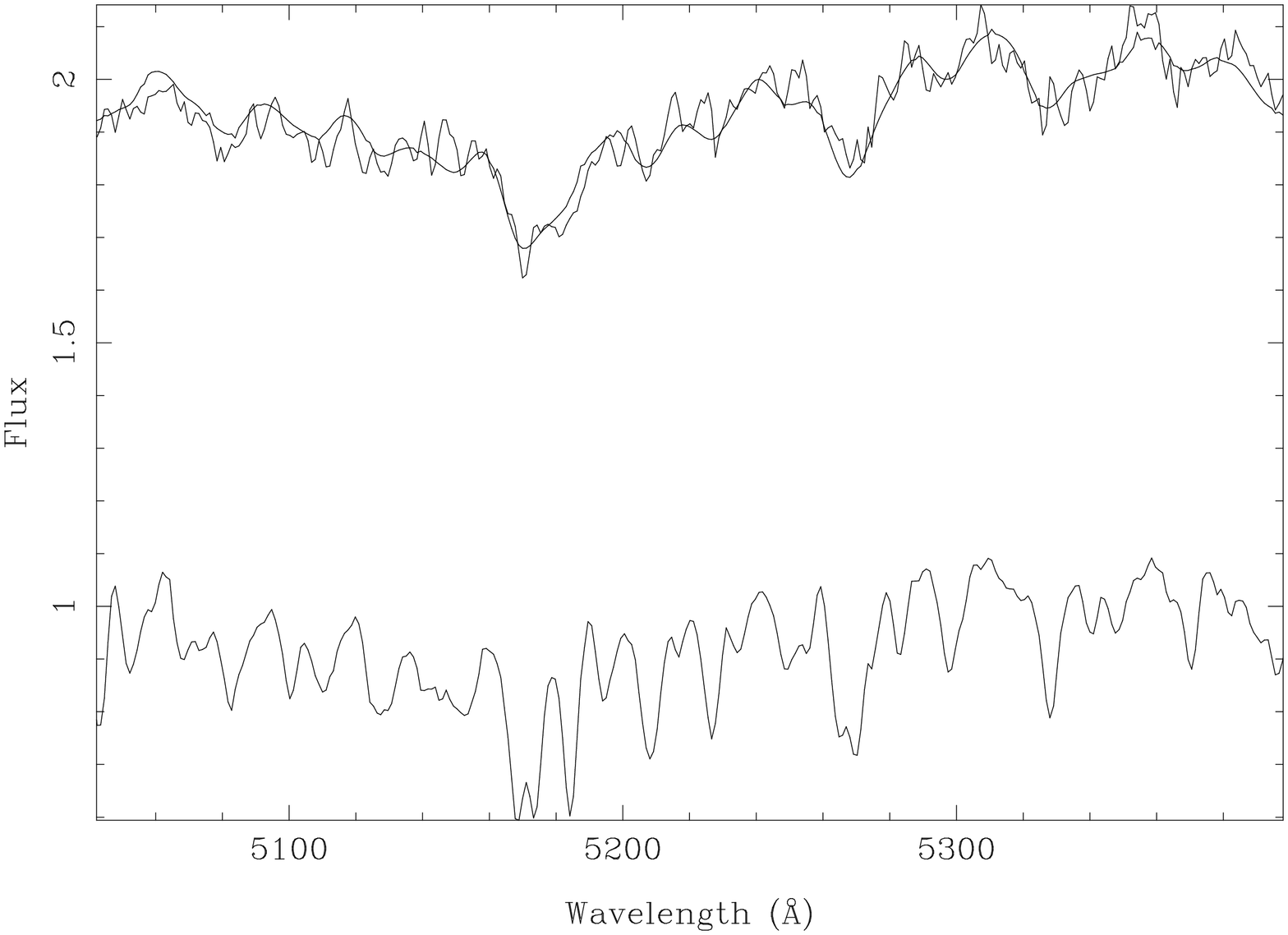,width=9.5cm,angle=-90}
\figcaption[spectra.ps] {Spectra for the template star (lower line),
the central 0.21\arcsec\ of NGC~3379 (noisy upper line), and the
template star convolved with the derived central LOSVD (smooth upper
line).\label{fig3}}
\vskip 0.3cm

The 68\% confidence band for the LOSVD is determined using a bootstrap
approach. We convolve the template star with the measured LOSVD to
provide the initial galaxy spectrum (the upper smooth line in Fig.~3);
from that initial galaxy spectrum, we then generate 100 realizations
and determine the LOSVD each time. Each realization contains randomly
chosen flux values at each wavelength position drawn from a Gaussian
distribution, with the mean given by the initial galaxy spectrum and
the standard deviation given by the root-mean-square of the initial
fit. The average root-mean-square for the continuum-divided spectrum
is 0.033. The 100 realizations of the LOSVDs provide a distribution of
values for every LOSVD bin, from which we estimate the confidence
bands.

\vskip 0.2cm
\hskip -0.8cm\psfig{file=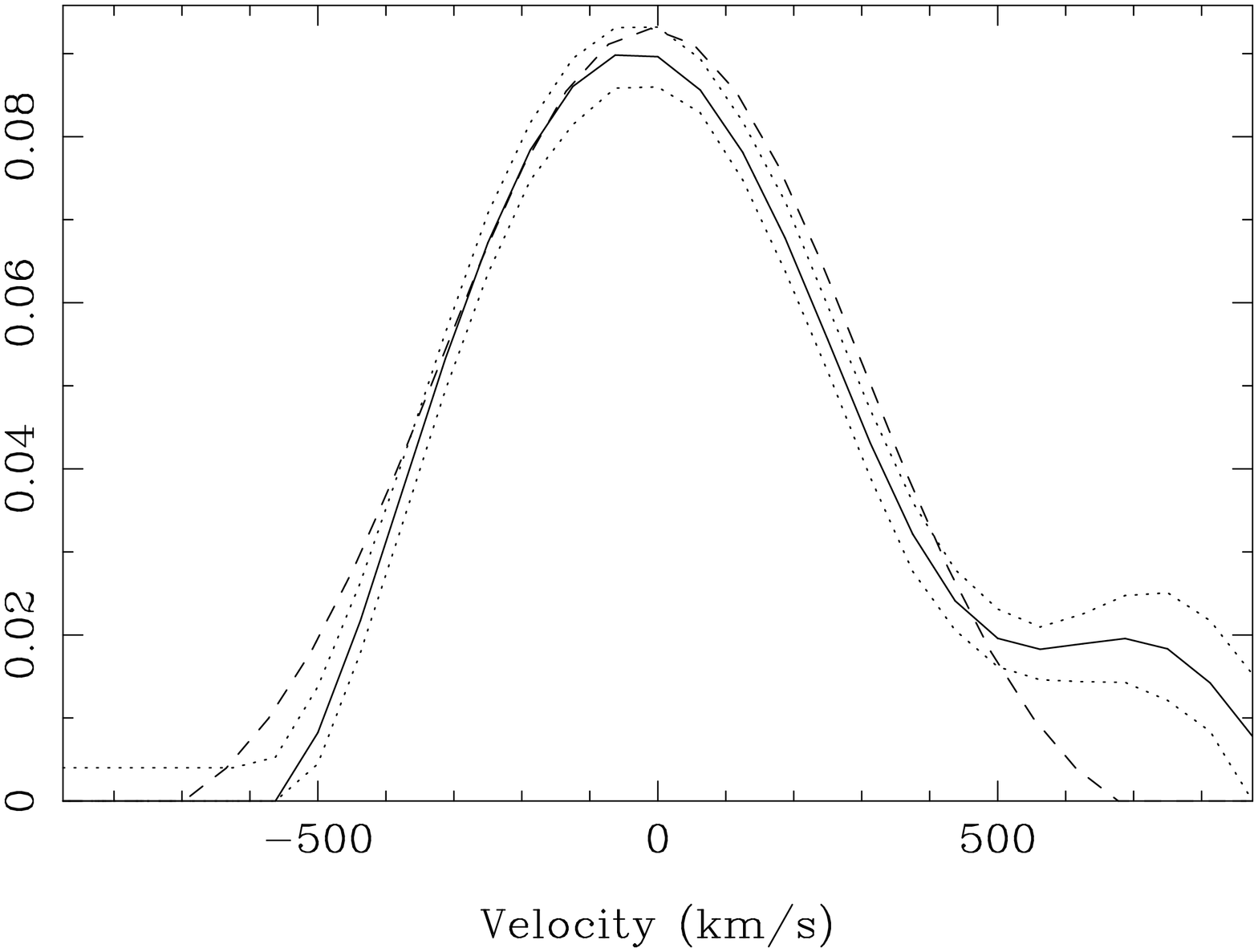,width=9.5cm,angle=-90}
\figcaption[losvd.ps]
{LOSVD for NGC~3379. The solid and dotted lines are the best
estimate and the 68\% confidence band from the central HST point. The
dashed line is a symmetrized LOSVD.\label{fig4}}
\vskip 0.3cm

\subsubsection{Asymmetry in the LOSVD}

Figure~4 presents the LOSVD with the 68\% confidence band (solid and
dotted lines). The LOSVD is significantly skewed with a tail to
positive velocities -- a feature also detected using the Fourier
correlation technique.  The skewness is crucial for the following
analysis and must be considered carefully. In an axisymmetric and even
a triaxial system, the {\it central} LOSVD is symmetric.

Various attempts were made to get rid of the skewness by reducing the
data differently; we used both templates individually, we used
flattened and un-flattened spectra, we used the Mg~b lines and the Na
doublet separately and combined, different shifts were used to combine
the spectra, and the spectra from both peak-ups were reduced
separately. In every case, the skewness remained. Since the skewness
appears to be real, we must conclude that the FOS was not centered,
that part of the galaxy light is being obscured, that the galaxy is
not symmetric near the center (possibly because of an offset nucleus),
or any combination of these affects. If the nucleus is offset, we
cannot successfully model NGC~3379 as an axisymmetric system. However,
there is no evidence to suggest that NGC~3379 has an offset nucleus
since the isophotes, except for the dust ring, are concentric.

Figure~5 is an image of the central 5\arcsec\ of the galaxy with the
spheroidal component subtracted. We include in Fig.~5 the exact
position of the FOS aperture. To determine the exact position of the
aperture, we have two independent checks: from the values used to
peak-up on the galaxy center, and from the acquisition image taken
immediately before the galaxy observations. For the peak-up procedure,
HST steps by 0.065\arcsec\ and points the FOS at the highest pixel
value. Examination of the peak-up numbers reveals that the likely
miscentering is 0.033\arcsec, since the center of the peak-up numbers
lies roughly between two bins in the peak-up array. This miscentering
is also confirmed from the acquisition image which places the FOS
aperture the same distance from the galaxy center.

\vskip 0.2cm
\hskip -0.8cm\psfig{file=image.ps,width=9.1cm,angle=-90}
\figcaption[image.ps]
{Image of NGC~3379 with the spheroidal light distribution
subtracted. The dust ring is clearly visible. The major axis of the
galaxy is shown as the solid line. The box is the size of the FOS
aperture (0.21\arcsec\ square). The uncertainty in the position of the
box is approximately one pixel (0.045\arcsec). The scale ranges from
$-200$ to $+100$ electrons per pixel.\label{fig5}}
\vskip 0.3cm

In order to explain the LOSVD skewness, the direction of the aperture
offset must correspond appropriately. HST placed the aperture on the
eastern side of the galaxy, along the major axis. Thus we acquire more
light from the eastern side relative to the western side. The NE side
of the galaxy has negative rotation velocities (Davies \& Birkinshaw
1988). If one side contributes more light and there is significant
rotation, the {\it mode} and mean of the distribution shifts towards
that side---in this case the NE side with negative velocity---and the
other side still contributes an extensive tail toward positive
velocities, thus generating the observed skewness in the LOSVD.

A further source for the LOSVD skewness may be that part of the galaxy
light is obscured. A dust ring is clearly visible, inclined by about
45\degr\ from the major axis (the solid line in Fig.~5). From the
relative depths of the obscuration on opposite sides, it appears that
the SW side of the ring is on the near side of the galaxy, while the
NE side is on the far side. This configuration means that the crucial
signal from stars near the nucleus is preferentially obscured on the
SW side, thus leading to the same effect as the miscentering; the
western side contributes less light than the eastern side causing
positive velocity skewness.

The miscentering of the aperture results from two effects, errors in
the peak-up procedure and errors due to the dust obscuration. To
calculate the miscentering from dust, we find the best center by using
the galaxy light beyond the dust ring and compare this with the FOS
center. HST points FOS to the highest position in a 0.21\arcsec\
aperture on a $4\times4$ raster with spacing of 0.065\arcsec. The dust
does not appear to have a significant effect on the location of the
brightest pixel; it causes a miscentering of only
0.01\arcsec. However, errors in the peak-up procedure are more
significant, and as mentioned above, the miscentering can be
0.033\arcsec. We conclude that the maximum miscentering is
0.04\arcsec\ (peak-up errors + dust miscentering). We then place the
FOS aperture in the WFPC2 image at a position 0.04\arcsec\ East from
the galaxy center and determine the relative amount of light in the
aperture from opposite sides of the galaxy. At this position, the
light from the NE side of the galaxy is twice as strong as from the SW
side. Thus, in the central bin of our model, we include all of the
light from the NE side and only half from the SW side.

An alternative is to use the best symmetrized LOSVD in the modeling
and assume equal contributions from both sides of the galaxy. The
dashed line in Fig.~4 is the maximum penalized likelihood estimate of
the best symmetric LOSVD. Results from both assumptions are discussed
below.

\subsection{Ground-Based Spectroscopy}

Ground-based spectroscopic data are available from Gebhardt~\&
Richstone (1999) (along four position angles), Statler~\& Smecker-Hane
(1999) (along four position angles), Bender~\etal\ (1994) (major axis
only), Gonzalez (1994) (major and minor axes), Franx~\etal\ (1989)
(major and minor axes), and Davies \& Birkinshaw (1988) (major, minor,
and 30\degr\ axes). The data from Franx~\etal\ and Davies~\&
Birkinshaw have lower S/N and will not be considered here. The data
from the first four sources all have high signal-to-noise (S/N) and in
general agree with one another (see Gebhardt~\& Richstone for a
detailed comparison), with the largest disagreement coming from the
large radii data. For the data of Gebhardt~\& Richstone, the position
angles observed and the radial extraction of the galaxy spectra were
designed to correspond exactly with the binning used in the modeling
(described in the next section), and we use their data.

\section{Models}

We use 3-integral axisymmetric models (Richstone~\etal\ 1998). The
models are orbit-based (Schwarzschild 1979) and constructed using
maximum entropy (Richstone \& Tremaine 1988). We run a representative
set of orbits in a specified potential, and determine the non-negative
weights of the orbits to best fit the available data. Maximizing the
entropy helps to provide a smooth phase-space distribution function.
Rix~\etal\ (1997) present a similar code which has been applied to the
spherical case; Cretton~\etal\ (1999) and \vdm~\etal\ (1998) developed
a fully-general axisymmetric code that is also orbit-based. The
difference between our code and theirs is due to how we impose
smoothness and how we fit to observational data. Smoothness in our
code comes from maximizing entropy, whereas the other authors enforce
smoothness of the distribution function directly by minimizing its
variation. The most important difference is that we fit the LOSVD
directly, and both Rix~\etal\ and Cretton~\etal\ fit moments of the
velocity profile. Using the LOSVD precludes the need to approximate
the velocity profile with a parametric function both in the
observations and in the orbital libraries.

\subsection{Generating Orbit Libraries}

The surface brightness yields an estimate of the luminosity density
distribution as explained above (\S 2.1), assuming that the luminosity
density is axisymmetric on similar spheroids and that the galaxy has a
given inclination. To obtain the mass density distribution, we have
made one of two assumptions about the mass-to-light ratio (\mtl):
({\it i}) constant \mtl, ({\it ii}) \mtl\ that varies according
to the $V-I$ color shown in Fig.~2. NGC~3379 is redder in the center
by $\Delta(V-I)~\simeq 0.1$. The analysis of Gonzalez (1994) indicates
that this color change is due mainly to metallicity (\ie,
no age effects). In this case, models of Worthey (1994) predict that
\mtl\ may be as much as 25\% higher in the center. The $V-I$ variation
in Fig.~2 is approximately log-linear, so we vary \mtl\ by 25\% from
the outer to inner radii following the same functional form. Finally,
we add a central point mass and compute the potential.

Using the derived potential, we follow orbits which sample the
available phase space in energy ($E$), angular momentum ($L_z$), and
the third integral ($I_3$) (Richstone~\etal\ 1998). Each orbit crosses
the equatorial plane 40 times. Binning the orbital distributions both
in real space and velocity, and in projected distance and
line-of-sight velocity, provides an estimate of the contribution to
the LOSVD from each orbit at each spatial bin. Both the real and
projected spatial binning are approximately logarithmic in radius and
linear in cos~$\theta$, where $\theta$ is measured from the pole. This
binning scheme provides approximately equal-mass bins.  Since we are
using axisymmetric modeling, we follow orbits only in $r$ and
$\theta$. We therefore have to choose the azimuthal angle for the
projection, and, since all angles are equally likely, we choose that
angle at random. For every time step, 100 azimuthal angles are chosen.

We measure an orbit's contribution on a grid of 80 radial bins, 20
angular bins, and 13 velocity bins---compressed to $20\times5\times13$
when comparing to the data. The size of the central bin equals the
aperture used in the FOS (0.21\arcsec). The outermost bin ranges from
149\arcsec\ to 200\arcsec. The centers of the five angular bins up
from the major axis are at 6\degr, 18\degr, 30\degr, 45\degr, and
72\degr. The velocity bins are about 100~\kms\ wide, approximately
equal to the spectral resolution of the ground-based observations.

We follow $\sim 3200$ orbits per model depending on inclination and
black hole mass. This number is only for one sign of the angular
momentum; consequently, it must be doubled before comparison to the
data. We flip the LOSVD at every position in the galaxy to provide the
appropriate reversed orbit. Thus, our final model contains $\sim 6400$
orbits. Varying the relative weights of these orbits allows us to
determine the best match to the data.

\vfill\eject
\subsection{Modeling the Data}

It is desirable to constrain every model bin with an observed
LOSVD. For NGC~3379, we have ground-based observations of the LOSVD
along four of the five position angles used in the modeling. In fact,
the observations were designed to mimic the model angular bins listed
above. Furthermore, the spectra were extracted from the data using the
same radial binning scheme as the modeling. Therefore, most model
bins correspond {\it exactly} to an observation; \ie, no interpolation
of the data or the model is necessary for comparison. This
correspondence breaks down for the ground-based data in the innermost
bins because of seeing.

The ground-based data extend out to 64\arcsec, about two effective
radii. The total number of bins for which we have an observation is
54. The model LOSVDs are calculated at 13 velocity positions with a
spacing of 100~\kms, providing at most 13 independent measurements of
the LOSVD per position. We therefore have a {\it maximum} of 702 (54
positions $\times$ 13 LOSVD bins) independent observables. However,
the actual number is smaller than this, mainly due to the smoothing
used in the measurement of the LOSVDs (discussed in \S5 and
Gebhardt~\& Richstone 1998), which correlates the 13 velocity
bins. Thus, we will overestimate the goodness-of-fit since we use the
68\% confidence bands of the LOSVDs to calculate it.

The model incorporates seeing by convolving the light distribution for
every orbit with the appropriate PSF before comparing to the data.
For the ground-based data, the estimated PSF, approximated as
Gaussian, has a FWHM of 1.5\arcsec, including both atmospheric seeing
and slit size. The HST data point was taken in a square aperture of
0.21\arcsec. The PSF for HST has a FWHM of around 0.07\arcsec, and
hence smearing due to the PSF is negligible compared to the aperture
used. Therefore, for the HST data, only the aperture size was
considered with no PSF convolution. We have checked in a number of
models that this approximation has no effect.

For each model, we match both the light in each of the $20\times5$
real-space bins, and the LOSVDs in the 54 locations where we have
data. The technique minimizes the $\chi^2$ between the model and data
LOSVDs. The light from the model matches the deconvolved spatial
brightness profile to a tolerance of better than 1\% in each bin,
which also ensures that the surface brightness matches better than 1\%
in any projected light bin. The match to the total light in each bin
has to be done using the real spatial profile and not the surface
brightness profile. If the matching were done in projection, it would
admit different density distributions than the one for which the
orbits were computed since the deprojection of a flattened, inclined,
axisymmetric system is not unique.

\section{Results}

\subsection{Estimating the Goodness-of-Fit}

For each model, the 702 observables (13 LOSVD bins at the 54
positions) determine the total $\chi^2$, given by $\sum_{i=1}^{702}
((y_i-y_i^{'})/\sigma_i)^2$ where the $y$'s are the LOSVD bin heights
for the model and the data and $\sigma$ is the uncertainty from the
68\% confidence band. As an example, Figure~6 plots the LOSVDs for
both the data and the best model at four positions: the center, a
major-axis position near to the center, a position close to one
effective radius on the major axis, and a minor-axis position near to
the center.

Judging the goodness-of-fit from the $\chi^2$ values of the various
models is not straightforward because the number of the degrees of
freedom is not easy to determine; the derivation of the LOSVDs
includes a smoothing parameter which correlates the values and
uncertainties of the velocity bin heights (Gebhardt~\& Richstone
1999). With higher S/N data than present, we could reduce the
smoothing parameter to lessen this problem.  We do not attempt to
estimate directly the actual degrees of freedom and therefore do not
have an overall goodness-of-fit measure, but instead we calculate the
change in $\chi^2$ as a function of the three variables -- black-hole
mass, inclination, and \mtl. The lowest $\chi^2$ value thus provides
the best model, with the uncertainties given by the classical
estimators for the $\Delta\chi^2$'s.

\vskip 0.2cm 
\hskip -0.6cm\psfig{file=good.ps,width=9.3cm,angle=-90}
\figcaption[good.ps] {Projected LOSVDs at four positions: the central
HST/FOS position (top-left), along the major axis near the center
(top-right), along the major axis near an effective radius
(bottom-left), and along the minor axis near the center
(bottom-right). The data are the open circles with their corresponding
error bars. The solid points are the model values. The area is
normalized to the total light in that bin.  This model is our
preferred model, with $1\times 10^8~\Msun$ black hole and inclination
90\degr. Note that the central HST LOSVD has been flipped about zero
relative to Fig.~4 in order to measure the model with respect to
positive rotational velocity.\label{fig6}}
\vskip 0.3cm

\begin{figure*}[b]
\figurenum{8}
\vskip 0.2cm
\centerline{\psfig{file=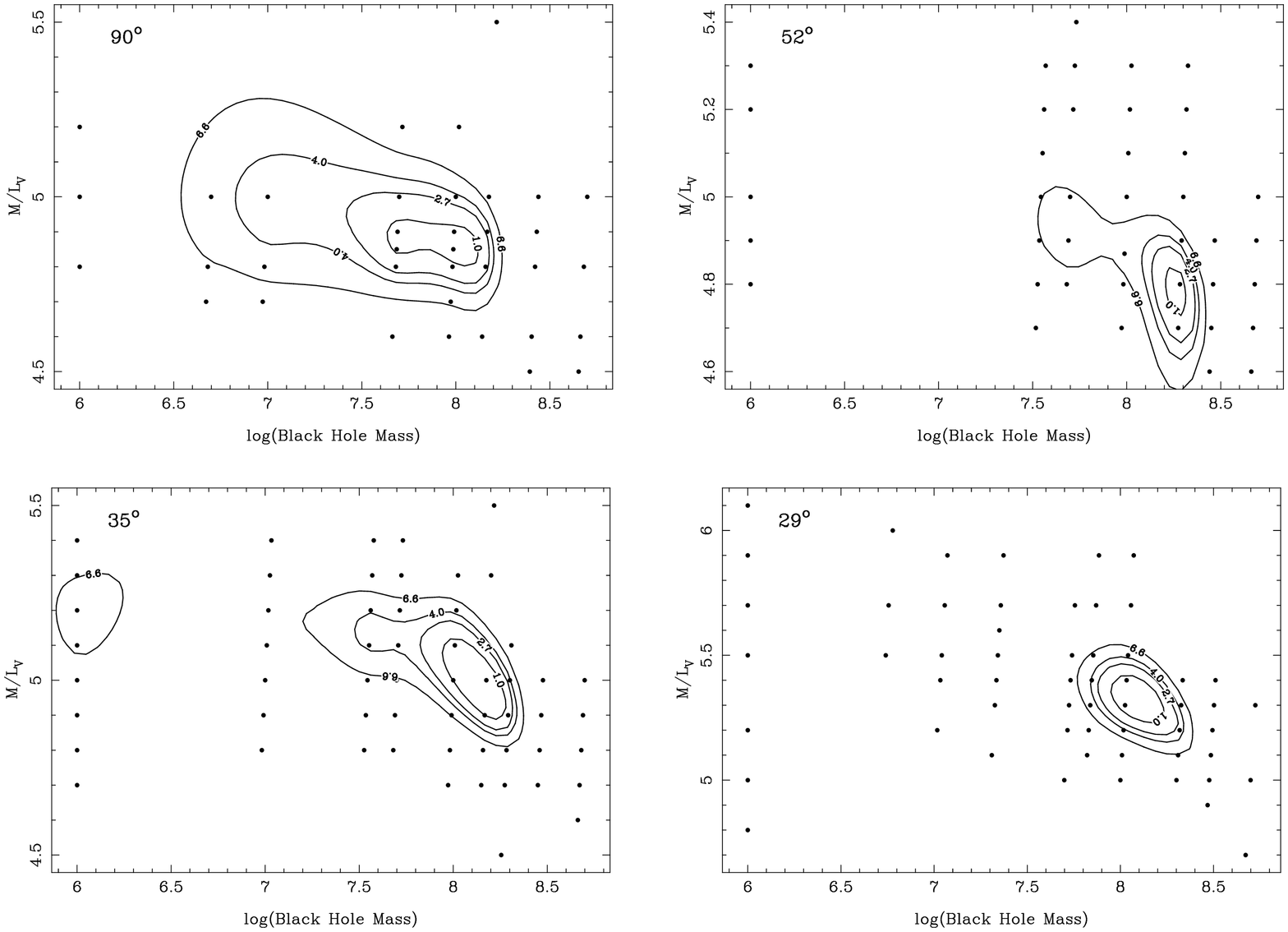,width=14.2cm,angle=90}}
\figcaption[comb.ps]
{Contours of $\Delta\chi^2$ as function of black-hole mass and
\mtl\ for four inclinations. The solid points represent the models that
were run. The contour values are equal to
$\Delta\chi^2 = 1,2.7,4.0, {\rm and}~6.6$; one-dimensional projection
of these contours provide the 68\%, 90\%, 95\% and 99\% confidence
bands respectively for the two variables, black-hole mass and \mtl.
\label{fig8}}
\end{figure*}
\vskip 0.3cm

The total $\chi^2$'s for most of the models have a value near 200,
giving a reduced $\chi^2$ of 0.3, assuming one observable equals one
degree of freedom. A reduced $\chi^2$ near unity would require either
that every three LOSVD bins are correlated by the smoothing, therefore
reducing the number of degrees of freedom by three, or that the
uncertainties of the bin heights are overestimated by a factor of
$\sqrt{3}$. In the former case, using $\Delta\chi^2$ as a
discriminator for the error of the best-fitting models will not be
affected since the number of degrees of freedom has no influence on
the calculation of the $\chi^2$ differences; $\Delta\chi^2 = 1$ will
still correspond to 1-$\sigma$. However, in the latter case, both the
$\chi^2$ and $\Delta\chi^2$ values would increase by a factor of
three, creating a smaller confidence band on the variables;
$\Delta\chi^2 = 1/3$ would then correspond to 1-$\sigma$. It is
likely, however, that the dominant reason for the low reduced $\chi^2$
is the underestimation of the degrees of freedom since we know {\it a
priori} that the smoothing must have an effect. To provide a quick
check on smoothing effects, we refit the LOSVDs for a subset of
positions using a much smaller smoothing length. As expected, the
LOSVD becomes noisier. Measuring the $\chi^2$ between this new LOSVD
and the model that we fit using the the older LOSVD, we find that the
increase in $\chi^2$ ranges from 3-5. Thus, we conclude that the
dominant effect of the smoothing is to correlate the variables and
reduce the number of degrees of freedom. In retrospect, it would have
been better not to have smoothed the LOSVDs and preserve their
statistical independence. Note however that our decision to accept
$\Delta\chi^2 = 1$ as the 1-$\sigma$ error is conservative and, if
anything, overestimates the errors.

\subsection{Model Fits}

We examined models with inclinations from 29\degr\ to 90\degr\
(intrinsic E5 to E1 galaxy), black holes from zero to $10^9\Msun$, and
\mtl\ from 4.0 to 8.0 in solar units. The $\chi^2$ is a function of
these three variables. For each inclination, Fig.~7 plots $\chi^2$ as
a function of black-hole mass (zero black-hole mass is shown as
$10^6$), using that \mtl\ which provides the smallest $\chi^2$ at that
black-hole mass. All inclinations show a minimum $\chi^2$ at
black-hole masses in the range $6\times10^7 - 2\times10^8\Msun$. The
fits clearly prefer the edge-on models (90\degr). Fig.~8 presents
contours of $\Delta\chi^2$ as a function of black-hole mass and \mtl\
for each inclination. The points represent locations of the modeled
values. The contours use a two dimensional smoothing spline (Wahba
1980) to estimate the $\chi^2$ values at positions where no models
occur. As in Wahba, Generalized Cross-Validation determines the
smoothing value; however, the modeled values are relatively smooth and
little smoothing is necessary.

\vskip 0.4cm
\hskip -1cm\psfig{file=plotall.ps,width=9.1cm,angle=-90}
\figcaption[plotall.ps]
{$\chi^2$ as function of black hole mass for various
inclinations. The labels give the inclination and the intrinsic galaxy
shape. We have added $10^6\Msun$ to the models with no black hole to
put them on the log scale.\label{fig7}}
\vskip 1.0cm

Since we have such a large parameter space---minimizing several
thousand variables---we must ensure that the modeling program is
finding a true minimum. Furthermore, we require that the numeric noise
caused by the minimizer be smaller than the quoted uncertainties of
the output parameters (e.g., the black hole mass). The two conditions
that govern these checks are the sampling densities that we use in
parameter space, and the tolerance used to determine convergence in
the minimizer. Fig.~8 demonstrates the sampling densities for the
black-hole mass and the \mtl, where each point represents one
model. Fig.~7 shows the range and number of modeled inclinations.  The
smoothness of the $\chi^2$ contours in both figures demonstrate that
we have found the global minimum. As a further check, we have run
models with both larger and smaller \mtl\ than those presented,
however the fits are significantly worse and we do not include them in
the plots. Thus, we conclude that the sampling density provides
adequate coverage around the global minimum. Our second concern is
whether the tolerance used for the stopping criteria in the minimizer
creates significant noise for the parameter estimation. We must use a
stopping tolerance in the program or the time it takes for the
$\chi^2$ to asymptote becomes impractically long. Both Figures~7 and 8
provide an estimate of the numeric noise. From the soothness of plots,
the change in $\chi^2$ between two adjacent points is much smaller
than the global change in $\chi^2$. Also, in a handful of models, we
iterate till there is no change in $\chi^2$ relative to machine
precision, and find that this asymptotic value is within
$\Delta\chi^2=0.02$ of the stopping value. Therefore, the noise in
$\chi^2$ generated from the minimization routine is not
significant. The smoothness is the solution is also seen in Fig.~9,
where we plot best-fit black hole mass as a function of
inclination. Although, the sampling of points is not dense, the fact
that there are no abrupt changes signifies we are not subject to
numeric noise in the solution.

\setcounter{figure}{8}
\vskip 0.2cm
\psfig{file=plotbest.ps,width=8.0cm,angle=-90}
\figcaption[plotbest.ps]
{The best-fit black-hole mass as a function of inclination.
\label{fig9}}
\vskip 0.4cm

\vskip 0.2cm
\psfig{file=int.ps,width=9.1cm,angle=0}
\figcaption[int.ps]
{Internal dynamics of NGC~3379 for the preferred model,
$1\times10^8~\Msun$ black hole, inclination 90\degr. The top plot
shows the three internal dispersions along the major axis; the bottom
plot shows the mean velocity in the equatorial plane.\label{fig10}}
\vskip 0.3cm

As described in \S3.1, we have also run models with \mtl\ increasing
25\% log-linearly with radius towards the galaxy center. In this case,
the required \mtl\ is systematically lower than when using the
constant \mtl\ assumption; however, the results for the required black
hole and inclination are unchanged.

\begin{figure*}[t]
\centerline{\psfig{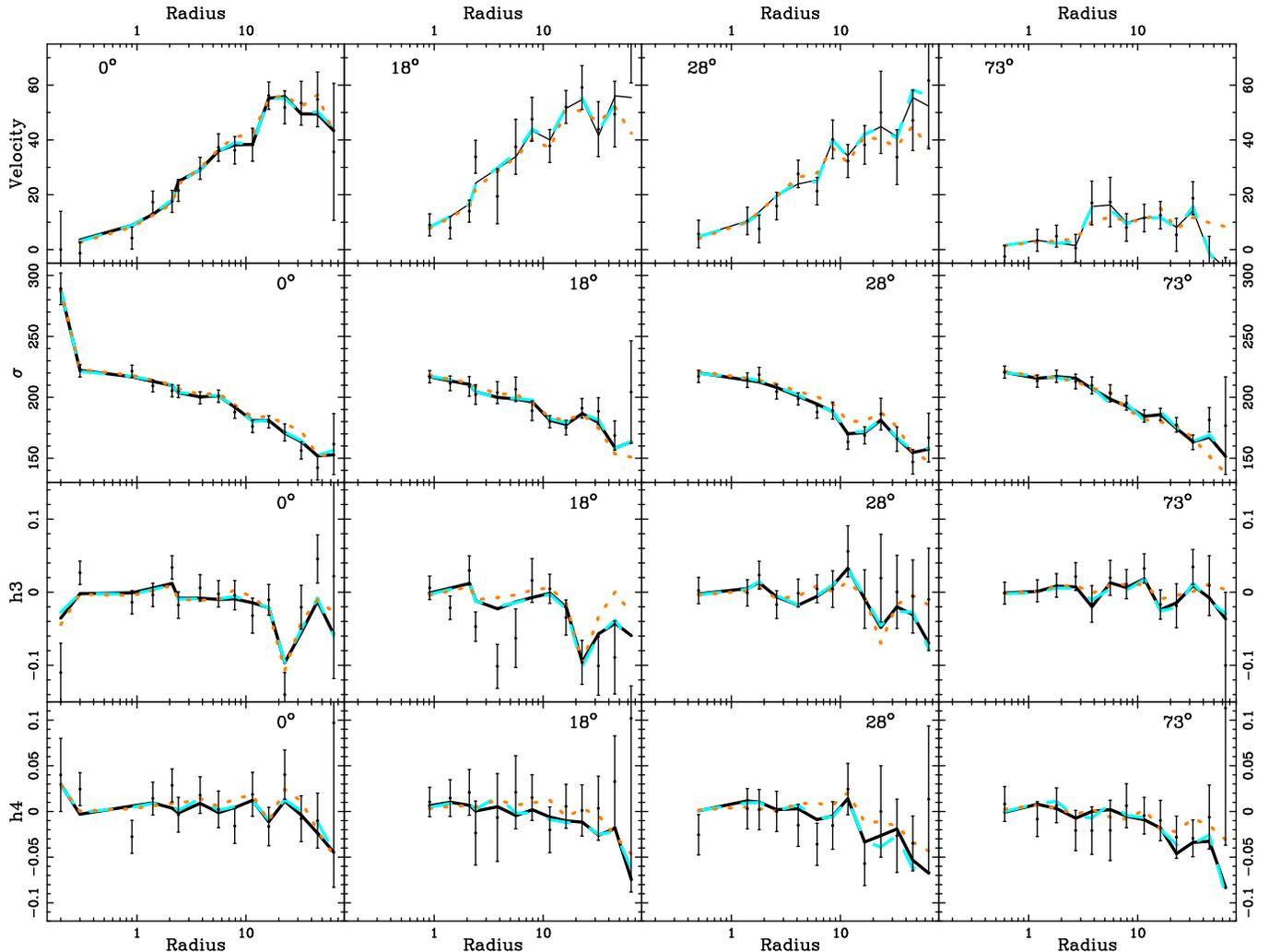}}
\figcaption[alldata.ps]{Gauss-Hermite moments for data---the points
with 68\% error bars---and models. The solid line is the best-fit
edge-on black-hole model, the dased blue line is the edge-on
no-black-hole model, and the dotted orange line is the best-fit
29\degr\ model.  Each column represents a different position angle
which is measured up from the major axis.
\label{fig11}}
\end{figure*}

The best model has an inclination of 90\degr, a black hole mass of
$1\times 10^8\Msun$, and \mtl$=4.85$. The allowed mass of black holes
range from $6\times10^7-2\times10^8\Msun$. The internal dynamics of
the best-fitting model are shown in Fig.~10. The top plot shows the
internal velocity dispersions of the three components along the major
axis. The model is tangentially biased (in the $\phi$ direction) in
the smallest bin, radially biased in the mid-range radii, and
tangentially biased again at the largest radii. Independent of
inclination, Fig.~10 well characterizes the orbital distribution for
each of the best-fitting models. Similarly, the mean velocity in the
equatorial plane (bottom plot of Fig.~10) has a general shape for the
best models; it shows no systematic trends from $0.4-30\arcsec$
(20-1500~pc), increasing significantly in the center due to the
presence of the massive black hole. Beyond about 60\arcsec\ the model
is unconstrained since there is no kinematic information, but we plot
the model results there since they represent the maximum entropy
configuration. For each inclination the trend with black-hole mass is
as follows: increasing the black-hole mass requires an increased
amount of tangential anisotropy in the central bin, a decrease of the
radial bias in the mid-range radii, and the orbits at the largest
radii to become more isotropic. The coupling between the black-hole
mass and the amount of radial anisotropy in the mid-radii stems from
the need to match the same central LOSVD; a large central dispersion
can be modeled with a large black-hole mass or radial
orbits. Therefore, as the black-hole mass increases, the need for
radial orbits decreases. In the very central bin, a large black-hole
mass induces large rotational velocities and also large dispersion in
the $\phi$ direction. At the largest radii, the situation is more
complex since we have not included a dark halo; if the projected
dispersion is larger than the dispersion obtained assuming isotropy
and the stellar mass profile then one can either invoke a dark halo or
include tangential anisotropy. In any event, it is not clear whether
meaningful analysis at large radii can be obtained from models which
do not include a dark halo.

To demonstrate why the data seem to prefer an edge-on model that
contains a black hole, we must look at the model fits to the
data. Figure~11 plots Gauss-Hermite polynomial estimates for the data
and model LOSVDs for all four position angles. The difference between
the edge-on model and the nearly face-on model is almost entirely at
large radii; the edge-on model matches the large-radial data better,
specifically for the velocity and velocity dispersion. In contrast,
the difference between the no-black-hole and black-hole model is so
subtle that one can barely discriminate those two models in
Fig.~11. Only in the central bin does the the black-hole model
slightly better estimate the skewness in the profile. However, Fig.~11
is not the optimal way to compare models since the fit uses the full
LOSVD and not moments. Instead of plotting all 54 LOSVDs for various
models, Fig.~12 plots the difference in $\chi^2$ at each spatial
position. This difference represents the $\chi^2$ measured for the
LOSVD of a particular model minus that of the best-fit model (edge-on,
black-hole model). The two comparison models are the same as in
Fig.~11, the best no-black-hole model (an edge-on model) and a nearly
face-on black-hole model (at 29\degr). As expected, the discrimination
between black-hole/no-black-hole comes mainly from the HST/FOS data
point. A no-black-hole model fails to match the shape of the central
LOSVD; however, there is a limit to size of the central black hole
since very high black-hole masses models over shoot the dispersions in
the central few ground-based measurements. Without the HST data point,
we would not have been able to determine whether a black hole exists
in NGC~3379. The discrimination for the inclination comes from the
data at larger radii; here, the edge-on model is much better able to
match the projected kinematics. As we discuss later, this inclination
constraint is uncertain due to our assumption of constant \mtl.

\vskip 0.2cm
\psfig{file=dchi.ps,width=9.1cm,angle=-90}
\figcaption[dchi.ps]
{The change in $\chi^2$ relative to the best-fitted
model. Each panel represents a different observed position angle.  The
solid line represents the $\chi^2$ difference for the edge-on
no-black-hole model, and the dashed line represents the 29\degr\ model
with a $10^8\Msun$ black hole.\label{fig12}}
\vskip 0.3cm

There is an uncertainty in the absolute wavelength calibration of the
HST spectrum. Since we do not assume that the central HST LOSVD is
symmetric, we must determine the velocity zeropoint of this profile
relative to the rest frame of the galaxy ($920\pm10$~\kms,
Borne~\etal\ 1994). Unfortunately, our original observing strategy did
not include an accurate wavelength calibration, since we were
concerned mainly with the central dispersion, and thus did not obtain
the necessary flanking arc-lamp spectra. We therefore tried a range of
velocity zeropoints, models with velocity offsets from $0-50$~\kms. As
the offset increases, the lower limit on the black-hole mass grows,
whereas the upper limit remains the same. We are not able to fit
acceptable models with offsets greater than 15~\kms. Therefore, our
allowed range of black hole masses is largest using the 0~\kms\
velocity offset. All of the figures contain the results using this
value.

\section{Discussion}

The models in this paper depend primarily on the input black-hole
mass, inclination, and M/L assumption. A difficulty is that different
combinations of these variables can have similar appearances in the
projected kinematics. For example, the inclination and the M/L
variation both affect the fit at large radii; one can easily see how
differently inclined models of the same flattened galaxy will have
markedly different projected dispersion profiles, and also the
dispersion profile is obviously influenced by the inclusion and shape
of a dark halo. Thus, by varying the shape and amount of dark halo one
may be able to mimic the effect of changing the inclination. Since we
have not explored a full range of halo models, we will not make strong
claims as to our ability to constrain the inclination of the
galaxy. However, the black hole mass is mainly driven by the data at
small radii, and primarily by the HST/FOS data point. Fig.~9
demonstrates this independence from the inclination, as our best-fit
black hole mass does not depend strongly on the inclination.  An even
more robust measurement is the orbital structure. As either the
inclination or the black-hole mass varies, we find a similar velocity
ellipsoid; at the smallest bin the velocity ellipsoid is mainly
tangential, becoming radial in mid-range radii, and tangential again
at the largest radii. Fig.~13 plots the radial motion relative to the
tangential motion for four models at different
inclinations. Determination of the inclination, black-hole mass, and
orbital structure is discussed in detail below.

\vskip 0.3cm 
\hskip -0.7cm\psfig{file=rtot.ps,width=9.1cm,angle=-90}
\figcaption[rtot.ps] {Radial relative to tangential dispersion as a
function of radius along the major axis for the best-fit models at the
specified inclination. The tangential dispersion is defined as
$\sigma_{\rm tangential}^2 = (\sigma_{\theta}^2 +
\sigma_{\phi}^2)/2$.\label{fig13}}
\vskip 0.3cm

The inclination of NGC~3379 has been the subject of some debate.
Capaccioli~\etal\ (1991) conclude that NGC~3379 is similar to NGC~3115
but nearly face-on, with some degree of triaxiality. Capaccioli base
this conclusion only on the surface brightness profile and use no
kinematic data. In contrast, Statler~(1994) concludes that nearly
face-on, triaxial models are inconsistent with the data at the 98\%
confidence limit, strongly contradicting Capaccioli's result. He
prefers less-inclined models. Statler has obtained more recent data
(Statler~\& Smecker-Hane, 1999) and a fuller analysis is in
preparation.

As seen in Sect.~4, our best models prefer an edge-on
configuration. Preference for edge-on models arises from two
observations; first, the dispersion profiles along different axes are
slightly different, and, in an axisymmetric case, this configuration
is not possible for a face-on system. The difference mainly arises
from the inability to match the large radial data, however the face-on
model also cannot match the shoulder in the dispersion profile at
10--20\arcsec\ (previously noted by Statler~\& Smecker-Hane, 1999);
the shoulder is more pronounced in the central position angles, and
the edge-one models match this configuration well (see
Fig.~11). Second, major-axis rotation exists which is not allowed for
a face-on configuration of an oblate axisymmetric system. Thus these
two restrictions are mainly a result of our axisymmetry assumption.

More general models can be constructed than those presented here. We
could in principle extend the modeling technique to include triaxial
distributions but then the parameter space becomes too large for us to
search (each axisymmetric model takes approximately 20 hours on an
Ultrasparc). Second, our assumption that the galaxy is spheroidal
could be relaxed.  Romanowsky~\& Kochanek (1997) present a technique
which provides a range of different density distributions which
similarly project. Third, we have assumed that \mtl\ is independent of
radius (\ie, there is no dark halo). Ciardullo~\etal\ (1993), using
radial velocity measurements of planetary nebulae, find no evidence
for a dark halo in NGC~3379 out to 3.5 effective radii (120\arcsec);
however their uncertainties are large and could, in fact, allow for
some dark halo (Tremblay~\etal\ 1995). In all of our models, the last
velocity dispersion measurement (at 64\arcsec) along each position
angle is higher than the model velocity dispersion, possibly
suggesting the need for an increase in \mtl. At one effective radius
(34\arcsec), there is no need for an increase in \mtl, which is
consistent with previous studies of other galaxies (Kormendy~\&
Westphal 1989, Kormendy~\& Richstone 1992, Kormendy~\etal\ 1997).
Thus, assumptions of axisymmetry, spheroidal distribution, and
constant \mtl\ all lead to uncertainties in the best-fit inclination.
Detailed modeling including dark halos has been carried out by
Rix~\etal\ (1998) for NGC~2434 and Saglia~\etal\ (1999) for NGC~1399,
but only as spherical systems. Only by allowing for axisymmetric or
triaxial shapes with a dark halo can we set more realistic limits on
inclination.

Even though we are not able to place limits on the inclination, we can
measure black-hole limits. The presence of a dark halo will have no
effect on the need for a central black hole, or on its mass. The need
for the black hole is directly a result of the shape of the central
LOSVD observed with HST. The large dispersion and the skewness drive
the models to require a black hole. We have also run models assuming a
symmetrized central LOSVD (the dashed line shown in Fig.~4). The
allowable range of black hole masses remains the same for the inclined
models in this case. However, for the edge-on case the model with no
black hole fits as well. The symmetrized LOSVD has a dispersion of
275~\kms, compared with 289~\kms\ for the actual LOSVD, and has less
tail weight (as can be seen in Fig.~4). These differences permit a
no-black-hole model. The central bin most influences the model; most
of the orbits ($>70$\% for inclined models and $>95$\% for edge-on
models) contribute light to this bin, and, furthermore, many have
their pericenter there. Therefore, the characteristics of the central
LOSVD, in particular the high-velocity components, have a strong
effect on orbits throughout the galaxy. In the edge-on case with the
symmetrized LOSVD, the orbits redistribute themselves in such a way as
to create a slightly larger central dispersion (275~\kms) without the
need for a black hole, and at the same time maintain a fairly flat
projected dispersion profile; \ie, at small radii, the orbits are
mainly radial, whereas at large radii they become primarily
tangential. However, the actual LOSVD warrants matching the skewness
and the high-velocity tails. With no black hole present, the model
must use stars at larger radii to create the high-velocity components,
thereby limiting its ability to redistribute the orbits at large radii
and match the kinematics there. With a black hole, the black hole
provides much of the high-velocity tail of the LOSVD for orbits
locally and gives the model more freedom to better match the
larger-radii data. The symmetrized and skewed LOSVDs are inconsistent
with each other with greater than 95\% probability, both using an
estimate from the binned values and using the Gauss-Hermite
moments. Given the arguments in \S2.2.1, we must include the skewness
of the central LOSVD and thus do not consider the symmetrized LOSVD
models further.

Relaxing the other two assumptions---axisymmetry and spheroidal
distribution--is unlikely to change the black-hole limits. The models
with no black holes all are extremely poor fits to the data;
therefore, even if we allow for the variation in deprojected densities
given by Romanowsky~\& Kochanek, the results are unlikely to
change. Of course, since the deprojection is unique for the edge-on
models, our results are general in the edge-on case. Triaxial models
will allow even greater freedom to match the data, however it is not
clear whether triaxiality is a viable configuration in the centers of
galaxies (Merritt\& Fridman 1996, Merritt 1999), and we do not
consider these models.

The position angle of the dust ring in Fig.~5 is 45\degr\ away from
the major axis of the galaxy. A dust ring in an axisymmetric
system with an offset position angle is only neutrally stable but can
be stabilized by a small amount of triaxiality. If the galaxy is
nearly face-on, then triaxiality is not necessary to stabilize the dust
ring; in the face-on configuration, since the dust ring is close to
edge-on (about 75\degr), it must be in an orbit close to polar, and
thus can have arbitrary azimuthal angle. Velocity information along
the dust ring could provide an important check on the enclosed mass
derived from our models.

Pastoriza~\etal\ (1999), using ground-based gas kinematics, assume
that the gas disk in NGC~3379 is nearly face-on. Their mass estimate
inside 1.3\arcsec\ is $7\times 10^9\Msun$, strongly contradicting our
result. However, their large mass is due mainly to their disk
inclination assumption. If one uses the inclination as measured from
the dust ring, then the enclosed mass closely equals our value of
$1\times 10^8\Msun$; only with HST can we precisely determine the
gaseous disk configuration and determine whether the gas is even in a
disk at all. The HST/FOS spectrum includes the [NII] emissiom,
although since it is a single pointing, we do not have the radial
information to extract a meaningful dynamical analysis from the
gas. However, the width of the line does provide some information on
the underlying mass. The measured $\sigma$ is 200\kms, in agreement
with the mass deduced here; using the mass from Pastoriza~\etal\ would
imply a $\sigma$ around 700\kms. To further understand the gas
kinematics, we must have spatial information.

Van~der~Marel~\etal\ (1990) apply 2-integral flattened models for
NGC~3379 using constant anisotropy. They find an acceptable fit with
an inclination of 60\degr\ and anisotropic orbits. They suffer from
the same problem as we do for the inclination estimate; without
inclusion of a dark halo it is not possible to obtain adequate
inclination constraints. Furthermore, since our axisymmetric models
are fully general while their models specify a form for the
distribution function, we cannot directly compare our results to
theirs. Van~der~Marel~\etal\ conclude that there is no evidence to
support the need for a third integral using data from the first and
second moments of the LOSVD from Davies \& Birkinshaw (1988). Results
presented here, which are based on higher S/N data and are determined
from the full LOSVD, show that the best model requires three integrals
and is not consistent with a 2-integral distribution function; the
internal dispersions along the long axis in the radial and $\theta$
directions are significantly different, in direct conflict with a
2-integral distribution function. As seen in Fig.~13, the ratio of the
$\theta$ to the radial dispersion in our best model have a radial
variation from 0.6 to 1.2, with an average around 0.75. For
comparison, this ratio is around 0.6 for the spheroid component of our
Galaxy using RR~Lyrae halo stars (Layden 1995).

In addition to the measured black-hole mass, the orbital structure
appears to be a robust feature as well. The radial anisotropy profile
along the major axis in Fig.~13 shows that this profile is nearly
independent of input inclination. For all acceptable models, the
velocity ellipsoid is tangentially biased in the central bin and
radially biased in the mid-range radii. These orbital properties are
generally true along any position angle. This central tangential bias
appears in other 3-integral models studied (Gebhardt~\etal\ 2000) and
may be a common feature of early-type galaxies. Tangential bias in the
central regions around black holes appear to be a common feature in
simulations (Quinlan~\etal\ 1995, Quinlan~\& Hernquist 1997, Merritt
\& Quinlan (1998), and Nakano~\& Makino 1999). The radial bias in the
mid-range radii has also been seen before; using 3-integral models,
both Gerhard~\etal\ (1999) for NGC~1600 and Saglia~\etal\ (1999) for
NGC~1399 find radial motion throughout the main body of the
galaxy. Thus, we may be beginning to measure common features of the
orbital properties in ellipticals---tangential motion in the central
regions and radial motion at mid-range radii. At large radii, we must
include dark halos since they make a significant contribution
there. The next step is to determine dominant evolutionary effects
from theoretical models and will require detailed comparisons with
N-body simulations.

Magorrian~\etal\ (1998), using the same ground-based data, fit
two-integral models to the observed second-order velocity moment
profiles and require a BH of $3.9^{+0.3}_{-0.4}\times10^8M_\odot$.
Our more general three-integral models would allow an even larger
range of BH masses had we similarly used only the LOSVD's second
moments.  When we fit to the full LOSVDs, however, their shape
requires that our models be mildly radially anisotropic, reducing the
required BH mass by a factor of at least two over Magorrian~\etal 's
result. The models presented in this paper still contain restrictive
assumptions---namely constant \mtl\ and spheroidal
symmetry---requiring yet more models to study black hole properties in
complete generality.

Kormendy (1993) and Kormendy \& Richstone (1995) find that there is a
correlation between the black hole mass and the mass of the
spheroid. Whether this relation is a real correlation or an upper
envelope that extends to smaller black-hole masses remains to be
seen. As the data quality and the modeling techniques improve, we
should be better able to constrain the relationship between black hole
and spheroid mass, and begin to measure correlations with the orbital
characteristics.

\acknowledgements

We thank Gary Bower, Richard Green, Luis Ho, and Jason Pinkney for
detailed discussions about the results. We thank an anonymous referee
who found an error in the previous version of this paper. We are
grateful to C.~D.~Keyes, J.~Christensen, and J.~Hayes for help with
the data analysis.  This work was supported by HST data analysis funds
through grant GO--02600.01--87A and by NSERC. KG is supported by NASA
through Hubble Fellowship grant HF-01090.01-97A awarded by the Space
Telescope Science Institute, which is operated by the Association of
the Universities for Research in Astronomy, Inc., for NASA under
contract NAS 5-26555.

\begin{deluxetable}{rrrrr}
\tablenum{1}
\tablewidth{20pc}
\tablecaption{NGC~3379 HST Photometry Results}
\tablehead{
\colhead{$R$ (\arcsec)}  & 
\colhead{$V$}            &
\colhead{$V-I$}          &
\colhead{$e$}            &
\colhead{PA}             }
\startdata
0.023   &  14.625  &  1.302  &  0.121  &  92.3  \nl
0.046   &  14.739  &  1.314  &  0.121  &  92.3  \nl
0.091   &  14.931  &  1.340  &  0.040  &  92.3  \nl
0.136   &  14.978  &  1.311  &  0.232  &  92.3  \nl
0.182   &  15.047  &  1.322  &  0.181  & 101.8  \nl
0.227   &  15.101  &  1.307  &  0.181  &  87.6  \nl
0.273   &  15.147  &  1.311  &  0.177  &  87.6  \nl
0.318   &  15.184  &  1.313  &  0.149  &  85.8  \nl
0.364   &  15.227  &  1.311  &  0.141  &  84.4  \nl
0.409   &  15.276  &  1.325  &  0.119  &  84.1  \nl
0.455   &  15.288  &  1.311  &  0.119  &  78.8  \nl
0.500   &  15.326  &  1.318  &  0.119  &  76.4  \nl
0.546   &  15.351  &  1.313  &  0.110  &  64.9  \nl
0.566   &  15.368  &  1.326  &  0.083  &  58.1  \nl
0.665   &  15.412  &  1.297  &  0.119  &  65.7  \nl
0.783   &  15.485  &  1.300  &  0.095  &  58.7  \nl
0.921   &  15.578  &  1.299  &  0.089  &  69.6  \nl
1.084   &  15.692  &  1.300  &  0.085  &  67.5  \nl
1.275   &  15.812  &  1.305  &  0.081  &  70.8  \nl
1.500   &  15.933  &  1.299  &  0.101  &  70.9  \nl
1.764   &  16.072  &  1.297  &  0.104  &  72.1  \nl
2.076   &  16.213  &  1.293  &  0.115  &  73.3  \nl
2.442   &  16.365  &  1.295  &  0.117  &  74.0  \nl
2.873   &  16.528  &  1.291  &  0.119  &  73.5  \nl
3.380   &  16.702  &  1.286  &  0.118  &  73.8  \nl
3.977   &  16.895  &  1.284  &  0.111  &  74.0  \nl
4.678   &  17.110  &  1.278  &  0.105  &  74.1  \nl
5.504   &  17.352  &  1.275  &  0.094  &  74.0  \nl
6.475   &  17.596  &  1.272  &  0.088  &  73.8  \nl
7.618   &  17.830  &  1.267  &  0.089  &  73.2  \nl
8.962   &  18.064  &  1.262  &  0.089  &  73.2  \nl
10.544  &  18.313  &  1.257  &  0.090  &  72.1  \nl
12.405  &  18.572  &  1.246  &  0.092  &  71.1  \nl
14.594  &  18.855  &  1.237  &  0.089  &  69.3  \nl
\enddata
\end{deluxetable}

\end{document}